\documentclass[prl,twocolumn,showpacs,superscriptaddress,floatfix]{revtex4}
\usepackage[latin1]{inputenc}
\usepackage{graphicx}
\usepackage{amsmath}
\usepackage{amssymb}
\usepackage{pifont}
\usepackage[dvips]{color}
\usepackage{subeqnarray}
\usepackage{psfrag}
\def\nablab{\mbox{\boldmath$\nabla$\unboldmath}}
\def\betab{\mbox{\boldmath$\beta$\unboldmath}}

\def\HS{\mbox{\scriptsize HS}}
\newcommand{\bb}{\mathbf{b}}

\newcommand{\kb}{\mathbf{k}}
\newcommand{\Ub}{\mathbf{U}}

\def\mean#1{\langle #1 \rangle}

\begin{document}

\title{ Effective slip over superhydrophobic surfaces in
  thin channels }

\author{Fran\c{c}ois Feuillebois}
\affiliation{CNRS UMR 7636 and 7083, ESPCI, 10 rue Vauquelin, 75005
Paris, France}

\author{Martin Z. Bazant} \affiliation{CNRS UMR 7636 and 7083, ESPCI,
  10 rue Vauquelin, 75005 Paris, France} \affiliation{Department of
  Mathematics, Massachusetts Institute of Technology, Cambridge, MA
  02139 USA} \affiliation{ Department of Mechanical Engineering,
  Stanford University, Stanford, CA 94305 USA}

\author{Olga I. Vinogradova}
\affiliation{CNRS UMR 7636 and 7083, ESPCI, 10 rue Vauquelin, 75005
Paris, France} \affiliation{A.N.~Frumkin Institute of Physical
Chemistry and Electrochemistry, Russian Academy of Sciences, 31
Leninsky Prospect, 119991 Moscow, Russia} \affiliation{ITMC and
DWI, RWTH Aachen, Pauwelsstr. 8, 52056 Aachen, Germany}

\date{\today}

\begin{abstract}
  Superhydrophobic surfaces reduce drag by combining hydrophobicity
  and roughness to trap gas bubbles in a micro- and nanoscopic
  texture. Recent work has focused on specific cases, such as striped
  grooves or arrays of pillars, with limited theoretical
  guidance. Here, we consider the experimentally relevant limit of
  thin channels and obtain rigorous bounds on the effective slip
  length for any two-component (e.g. low-slip and high-slip) texture
  with given area fractions. Among all anisotropic textures, parallel
  stripes attain the largest (or smallest) possible slip in a
  straight, thin channel for parallel (or perpendicular) orientation
  with respect to the mean flow. For isotropic (e.g. chessboard or
  random) textures, the Hashin-Strikman conditions further constrain
  the effective slip. These results provide a framework for the
  rational design of superhydrophobic surfaces.
      \end{abstract}


\maketitle

{\bf Introduction.}-- The design and fabrication of micro- and
nanotextured surfaces have received much attention in recent
years.  It has also been recognized that a modified surface
profile can induce novel wetting properties of a solid, which
could not be achieved without roughness~\cite{quere.d:2005}.
Depending on interfacial characteristics, the Wenzel state, where
the liquid impregnates the surface, can enhance wettability, or
the Cassie state, where the texture is filled with gas, can
dramatically amplify hydrophobicity ~\cite{bico.j:2002}. The
remarkable mobility of liquids on such superhydrophobic surfaces
renders them ``self-cleaning'' and causes droplets to roll (rather
than slide) under gravity and rebound (rather than spread) upon
impact. Beyond their fundamental interest, superhydrophobic
surfaces may revolutionize
microfluidics~\cite{stone2004,squires2005}, by reducing viscous
drag in very thin channels and amplifying transport
phenomena~\cite{ajdari2006} and transverse
flows~\cite{stroock2002a}.

Reduced wall friction is associated with the breakdown of the no-slip
hypothesis. It has recently become clear that liquid slippage occurs
at smooth hydrophobic surfaces, as described by the Navier boundary
condition~\cite{vinogradova.oi:1999,lauga2007,bocquet2007}
$v_s = b \partial v / \partial z,$
where $v_s$ is the slip (tangential) velocity at the wall and the axis
$z$ is normal to the surface. A mechanism for dramatic friction
reduction involves a lubricating gas layer of thickness $\delta$ with
viscosity $\mu_g$ much smaller than that of the liquid
$\mu$~\cite{vinogradova.oi:19951}, so that $b \approx \delta
(\mu/\mu_g - 1) \approx 50 \delta$~\cite{note0}. This scenario allows
to achieve slip length of only of a few tens of nm in case of smooth
hydrophobic surfaces~\cite{vinogradova.oi:2003}.  The presence of a
rough texture however stabilizes the gas layer, and by increasing its
height $\delta$, the slip length may reach tens of $\mu$m over the gas
regions. The composite nature of the texture, however, requires
regions of lower slip (or no slip) in direct contact with the liquid,
so the effective slip length of the surface $b^{\ast}$ (defined below)
is reduced. For anisotropic textures $b^{\ast}$ depends on the flow
direction and is generally a tensor~\cite{tensor}. Indeed,
experimental studies of flow past superhydrophobic surfaces suggest
that $b^\ast$ does not exceed several $\mu$m~\cite{joseph.p:2006}
and varies with the orientation of the wall texture relative to
flow~\cite{ou.j:2005}.

\begin{figure}
\psfrag{h}{$h$} \psfrag{L}{$L$} \psfrag{<U>}{$\langle U \rangle$}
\psfrag{<gradp>}{$\langle \nabla p \rangle$} \psfrag{b_1}{$b_1$}
\psfrag{b_2}{$b_2$}
 \includegraphics*[width=0.45\textwidth]{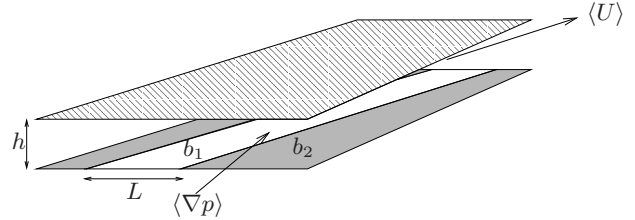}
\caption{Sketch of a thin channel, where the gap width $h$ is
small compared with the texture characteristic length $L$.}
\label{narrow_channel}
\end{figure}

The quantitative understanding of liquid slippage past
superhydrophobic surfaces is still challenging, and little theoretical
guidance is available for the design of optimal textures. Some exact
solutions are known for a flow on alternating (parallel or transverse)
no-slip and perfect slip
stripes~\cite{lauga.e:2003,wang2003,sbragaglia.m:2007} or transverse
inhomogeneous slip sectors~\cite{alexeyev:96}.  Simplified scaling
expressions have been proposed for a geometry of
pillars~\cite{ybert2007,bocquet2007}, and numerical approaches have
also been followed~\cite{cottin.c:2004,priezjev.nv:2005,benzi.r:2006}.
Nevertheless, general principles to maximize or minimize the effective
slip have not yet been established, even in the simple (but
experimentally relevant) lubrication limit, where the implication of
slip is the most pronounced~\cite{vinogradova.oi:19951}.

In this Letter, we propose a systematic approach to optimize the
effective slip length of a superhydrophobic surface in a thin channel,
based on the theory of heterogeneous porous
materials~\cite{Markov:2000,Torquato:2002}. We derive rigorous bounds
on the effective slip length for arbitrary anisotropic or isotropic
textures, depending only on the area fractions and local (any) slip
lengths of the high-slip and low-slip regions. In some cases, the
bounds are close enough to render detailed calculations unnecessary,
and in others the theory provides optimal textures which attain the
bounds (notably the maximum possible effective slip). Our theory also
predicts $b^\ast$ in certain geometries without requiring any
calculation.

{\bf Model and analysis.}-- We consider pressure-driven flow of a
viscous fluid between two textured parallel plates (``+'' and
``-'') separated by $h$, as sketched in Fig.\ref{narrow_channel}.
Motivated by superhydrophobic surfaces in the Cassie state, we
assume flat interfaces (such as idealization has been used in most of previous studies~\cite{lauga.e:2003,cottin.c:2004} and corresponds to a minimum dissipation in the system~\cite{sbragaglia.m:2007,hyvaluoma.j:2008})  characterized by spatially varying slip
lengths $b^+(x,y)$ and $b^-(x,y)$.
Our analysis is based on the
lubrication (or Hele-Shaw) limit of a thin channel, where the
texture varies over a scale $L \gg h$; the flow profile is then
locally parabolic at any position.

To evaluate the effective slip length, we calculate the velocity
profile and integrate it across the channel to obtain the
depth-averaged velocity $\Ub$ in terms of the pressure gradient
$\nablab p$ along the plates. As usual for the Hele-Shaw cell, the
result may be written as a Darcy law
\begin{equation}
 \Ub = - \frac{k(x,y)}{\mu} \nablab p ,
\label{Darcy}
\end{equation}
where we obtain the permeability
\begin{equation*}
 k(x,y) = \frac{h^2}{12} \left( 1 + \frac{3 (\beta^+
+\beta^-
+4 \beta^+ \beta^-)}{1+\beta^+
+\beta^-} \right)
\end{equation*}
in terms of the normalized slip lengths $\beta^+=b^+(x,y)/h$ and
$\beta^-=b^-(x,y)/h$. The permeability is maximized with two equal
surfaces, $\beta^+=\beta^-=\beta(x,y)$, so we consider this case (II)
with the goal of minimizing drag. We also consider the case (I) of one
no-slip wall ($\beta^+=\beta(x,y); \beta^-=0$), which is relevant for
various setups, where the alignment of opposite textures is
inconvenient or difficult. The permeability then takes the form:
\begin{equation}
 k(x,y) = \frac{h^2}{12}
\begin{cases}
1 + 3\beta(x,y)/[1+\beta(x,y)]  & \mbox{ case (I)}\\
 1 + 6 \beta(x,y)      & \mbox{ case (II)}
\end{cases}
\label{K}
\end{equation}
In general, the slip length may also vary locally with orientation, so
that $b(x,y)$ becomes a second-rank tensor $\bb(x,y)$, from which a
tensorial permeability $\kb(x,y)$ can be derived ~\cite{tensor}.

The slip length $b(x,y)$ (or $\bb(x,y)$) varies on the microscale
$L\gg h$, but we are interested in properties of the flow at the
macroscale. A natural definition of the effective slip length is based
on a hypothetical uniform channel with the same effective permeability.
First, we average (\ref{Darcy}) over the texture (denoted by
$\mean{.}$) at a mesoscale that is smaller than the macroscale, but
much larger than $L$, to obtain
\[
\mean{\Ub} = - \frac{1}{\mu} \mean{k(x,y) \nablab p} = -
\frac{\kb^*}{\mu} \cdot \mean{\nablab p}
\]
where in the last step we introduce the effective permeability
$\kb^\ast$, which is generally a tensor, even if $k(x,y)$ is locally
isotropic. Only with an isotropic structure at the mesoscale does it
become a scalar $k^*$. This definition is subject to the boundary
condition of a uniform pressure gradient $\nablab P$ applied at the
macroscale, which must equal the average pressure gradient,
$\mean{\nablab p}=\nablab P$, since the pressure is harmonic
\cite{Markov:2000}.

By analogy with (\ref{K}), we define the effective slip length in
terms of the effective permeability:
\begin{equation}
 k^*_j = \frac{h^2}{12}
\begin{cases}
1 + 3\beta^*_j/[1+\beta^*_j]       & \mbox{ case (I)}\\
1 + 6 \beta^*_j  & \mbox{ case (II)}
\end{cases}
\label{def_b_effective}
\end{equation}
where the principal (eigen)directions $j=1,2$ of $\kb^*$ correspond
with those of $\betab^*=\bb^*/h$, where $\bb^*$ is the effective slip
length tensor~\cite{tensor}.

Motivated again by superhydrophobic surfaces in the Cassie state, we
assume $b(x,y)$ switches between two values, $b_1$ and $b_2$,
associated with permeabilities $k_1, k_2$ from (\ref{K}), for regions
(or ``phases'') of liquid-solid and liquid-gas interfaces,
respectively. Let $\phi_1$ and $\phi_2$ be the area fractions of the
two phases with $\phi_1+\phi_2=1$.  We make no further assumptions in
deriving bounds on the effective slip length $\beta^*$ in a principal
direction (without transverse flow), aside from distinguishing between
anisotopic and isotropic textures.

{\bf Anisotropic textures.}-- In the general case of an
orientation-dependent texture ($\kb^* \ne k^* {\bf I}$), the Wiener
bounds apply for the effective permeability in a given
direction~\cite{Torquato:2002}: $ k^\perp \le k^* \le k^\parallel $.
The lower bound $k^\perp$ can be attained by parallel stripes
perpendicular to the pressure gradient: $
k^\perp = \left( \phi_1/k_1+ \phi_2/k_2 \right)^{-1} $.  The bound
$k^\parallel$ can also be attained, by stripes parallel to the
pressure gradient: $ k^\parallel = \phi_1 k_1 + \phi_2
k_2$. Physically, these special textures act like resistors in series
and in parallel, respectively.

\begin{figure}
\begin{center}
(a) \psfrag{parallel}{{\scriptsize$\parallel$}}
\includegraphics*[width=0.45\textwidth]{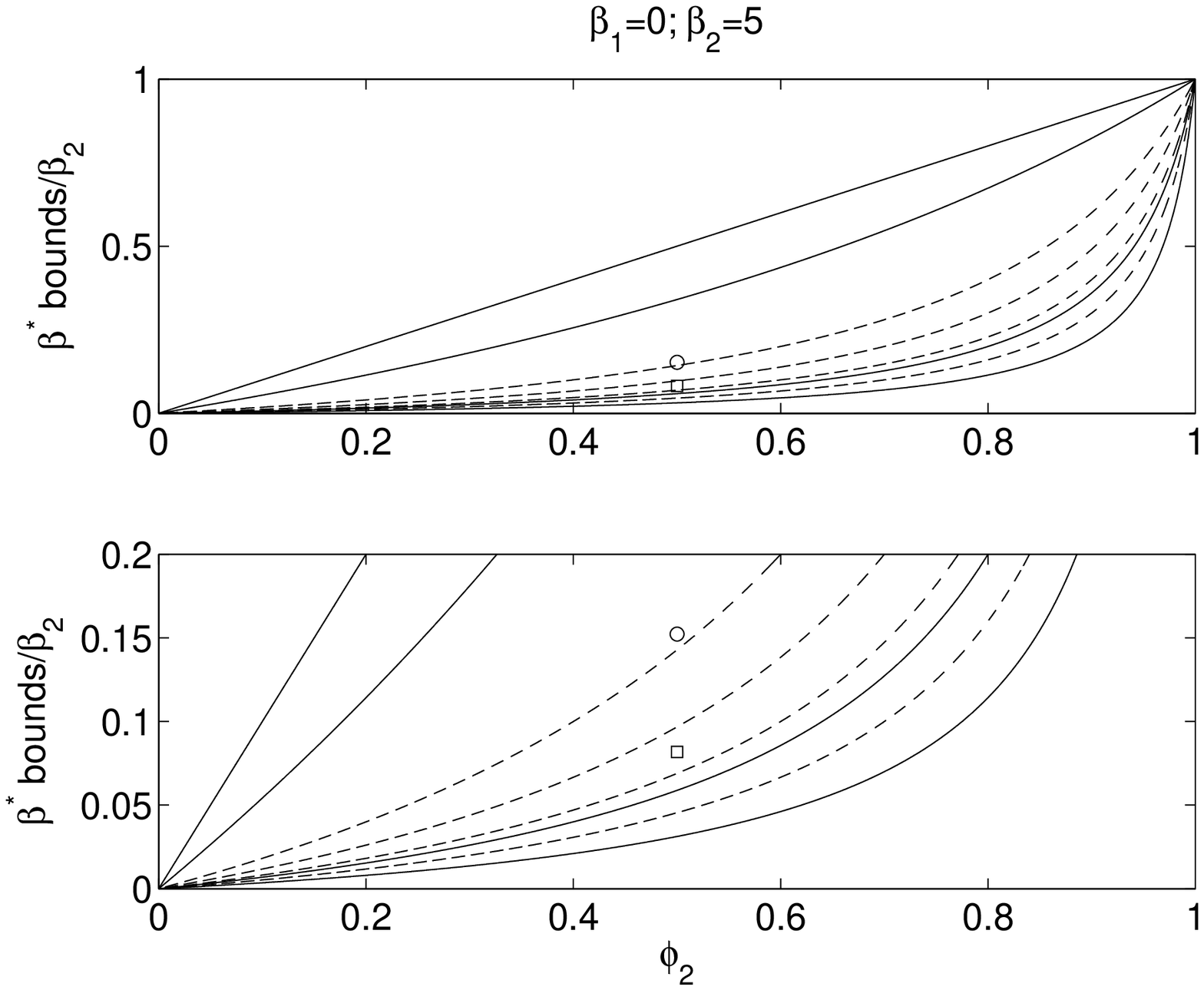}\\
(b) \psfrag{parallel}{{\scriptsize$\parallel$}}
\includegraphics*[width=0.45\textwidth]{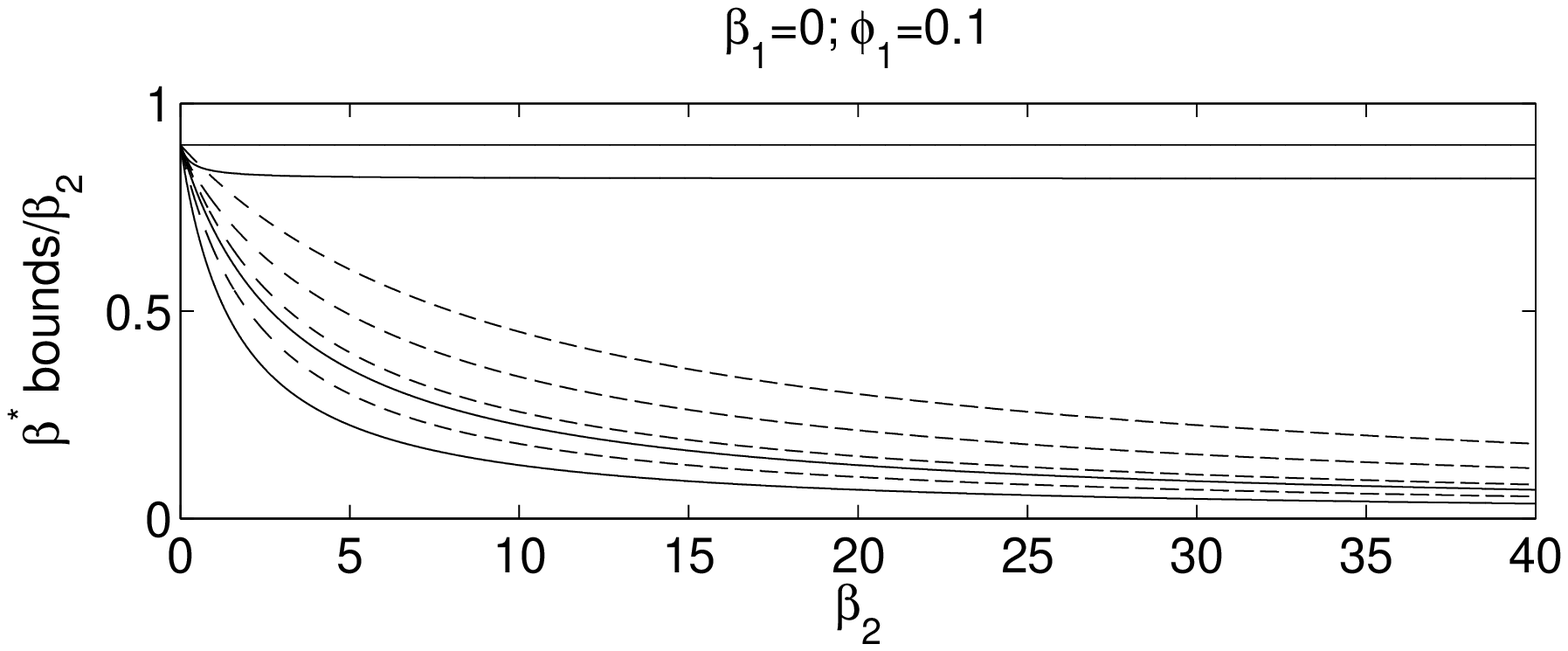}
\end{center}
\caption{(a) Bounds on the (normalized) superhydrophobic slip length
  $\beta^*/\beta_2$ versus the liquid-gas area fraction $\phi_2$,
  assuming no slip $\beta_1=0$ and high-slip $\beta_2=5$ on the
  liquid-solid and liquid-gas interfaces, respectively. Bottom: zoom
  of top figure. Dashed and solid lines correspond to cases (I) and
  (II) or one or two superhydrophobic surfaces, respectively.  In each
  case, curves from top to bottom represent: the upper bound for
  anisotropic, upper bound for isotropic, lower bound for isotropic,
  lower bound for anisotropic textures.  The value of $\beta^*$ for
  the chessboard or the isotropic Schulgasser structure sketched in
  Fig.~\ref{fig_designs} is also shown (square: case (I), circle: case
  (II)). (b) The same bounds plotted versus the slip length $\beta_2$
  for $\phi_2=0.9$. }
\label{fig_bounds}
\end{figure}

Using (\ref{K}) and (\ref{def_b_effective}), the corresponding bounds
for the effective slip length are
\begin{subeqnarray}
\frac{ \mean{\beta} + 4 \beta_1\beta_2}{1 + 4 \mean{\tilde{\beta}}}
\leq  \beta^*   \leq
\frac{ \mean{\beta} +  \beta_1\beta_2}{1 +  \mean{\tilde{\beta}}}
 & \mbox{ case (I)} \\
\frac{ \mean{\beta} + 6 \beta_1\beta_2}{1 + 6 \mean{\tilde{\beta}}}
\leq \beta^*  \leq
\mean{\beta}
 & \mbox{ case (II)}
\label{beta_Wiener_bounds}
\end{subeqnarray}
where
\begin{equation}
  \mean{\beta} = \phi_1\beta_1+\phi_2\beta_2 \ \mbox{ and } \
  \mean{\tilde{\beta}} = \phi_2\beta_1+\phi_1\beta_2
\label{def_mean}
\end{equation}
are the average slip length and average transposed slip length,
respectively. Using parameters for typical superhydrophobic surfaces,
these bounds are plotted versus the liquid-gas area fraction $\phi_2$
in Fig.~\ref{fig_bounds}(a) and versus the liquid-gas slip length
$\beta_2$ in Fig.~\ref{fig_bounds}(b).  In case (I) the bounds are
fairly close (especially when $\beta_2$ is large), so the theory
provides a good sense of the possible effective slip of any texture,
based only on the area fractions and local slip lengths. In case (II)
the difference between the upper and lower bounds is larger and grows
quickly with $\beta_2$. In either case, however, the texture attaining
the upper (lower) bound corresponds to stripes oriented parallel
(transverse) to the pressure gradient~\cite{note2}.

{\bf Isotropic textures.}-- Consider now any isotropic structure,
without a preferred direction ($\kb^* = k^* {\bf I}$). If the only
knowledge about the two-phase texture is $\phi_1, \phi_2$, then
the Hashin-Shtrikman (HS) bounds apply for the effective
permeability, $ k_{\HS}^L \le k^* \le k_{\HS}^U $, where (assuming
$\beta_1 \le \beta_2$ without loss of generality):
\begin{equation*}
  k_{\HS}^L = \mean{k} -\frac{\phi_1\phi_2[k]^2}{\mean{\tilde{k}}+k_1} ,
\qquad
 k_{\HS}^U = \mean{k} -\frac{\phi_1\phi_2[k]^2}{\mean{\tilde{k}}+k_2}
\end{equation*}
with $[k]=k_2-k_1$ and using the same notation as in
(\ref{def_mean}). Using (\ref{K}) and (\ref{def_b_effective}), the
corresponding bounds for the effective slip length are obtained in a
form similar to (\ref{beta_Wiener_bounds}):
\begin{equation}
\frac{\mean{\beta}+f(\beta_1)\beta_1\beta_2}
{1+f(\beta_1)\mean{\tilde{\beta}}}
\leq \beta^* \leq
\frac{\mean{\beta}+f(\beta_2)\beta_1\beta_2}
{1+f(\beta_2)\mean{\tilde{\beta}}}   \label{HS_bounds}
\end{equation}
where
\begin{equation}
f(\beta)=
\begin{cases}
(5+3\beta)/(2+5\beta) & \mbox { case (I) }\\
 3/(1+3\beta) & \mbox{  case (II) }
\end{cases}.
\end{equation}
The HS bounds (\ref{HS_bounds}) are plotted in Fig.~\ref{fig_bounds}
in the same way as the Wiener bounds (\ref{beta_Wiener_bounds}) and
behave similarly, aside from being closer and confined between them.
However, it turns out that isotropy does not dramatically reduce
(enhance) the maximum (minimum) effective slip in a thin channel,
especially in the configuration with two superhydrophobic surfaces,
case (II).

\begin{figure}
\begin{center}
(a)\includegraphics*[width=0.11\textwidth]{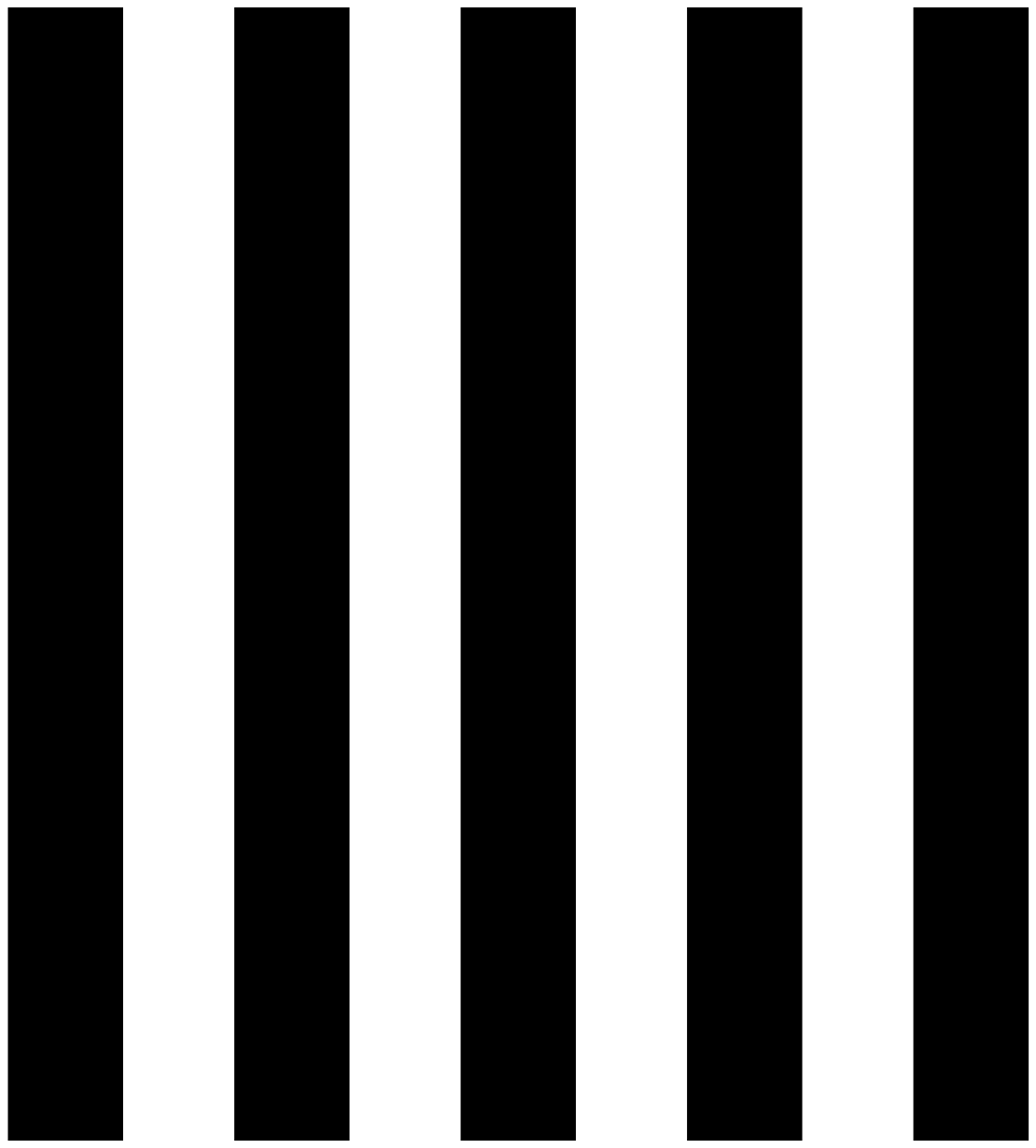}
(b)\includegraphics*[width=0.11\textwidth]{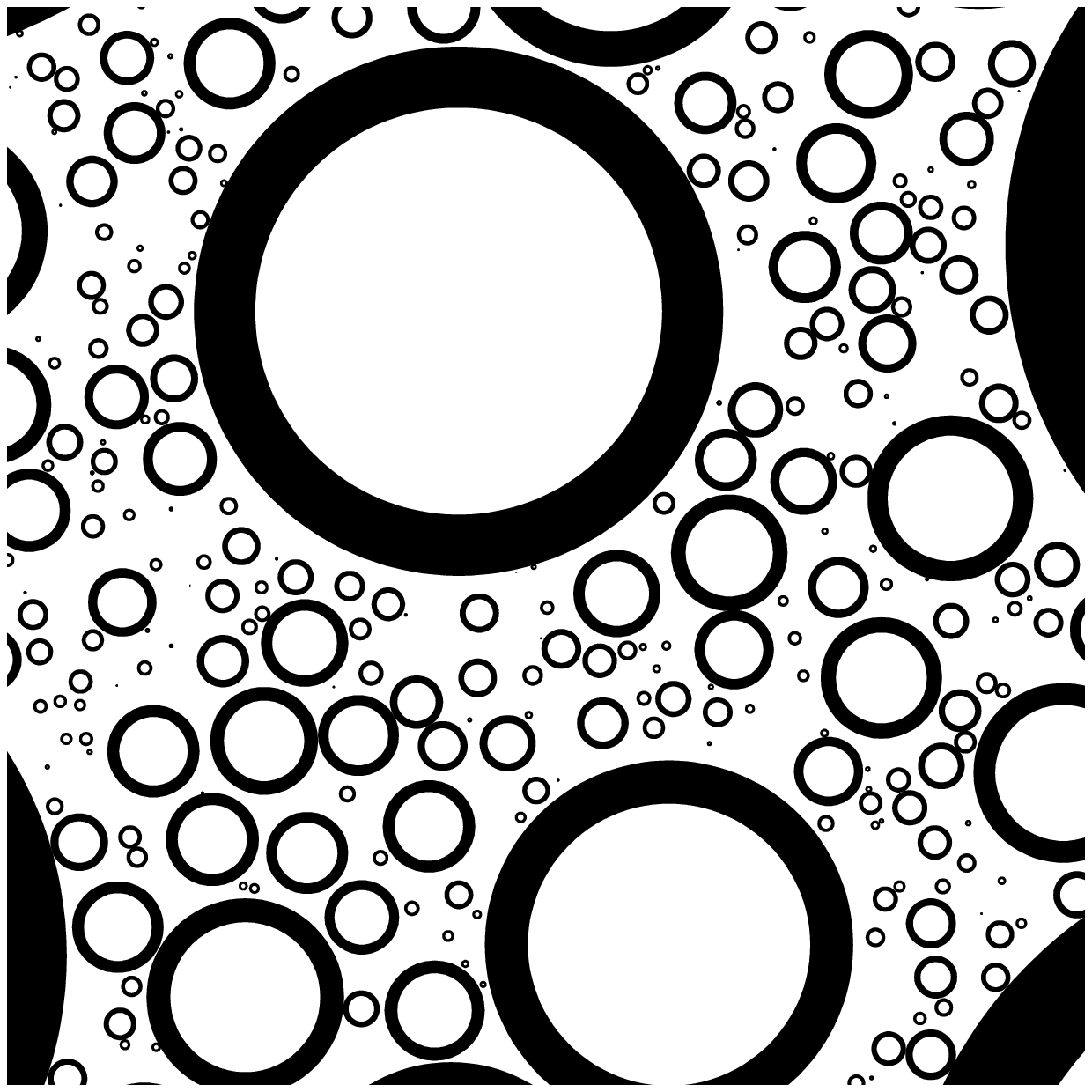}
(c)\includegraphics*[width=0.11\textwidth]{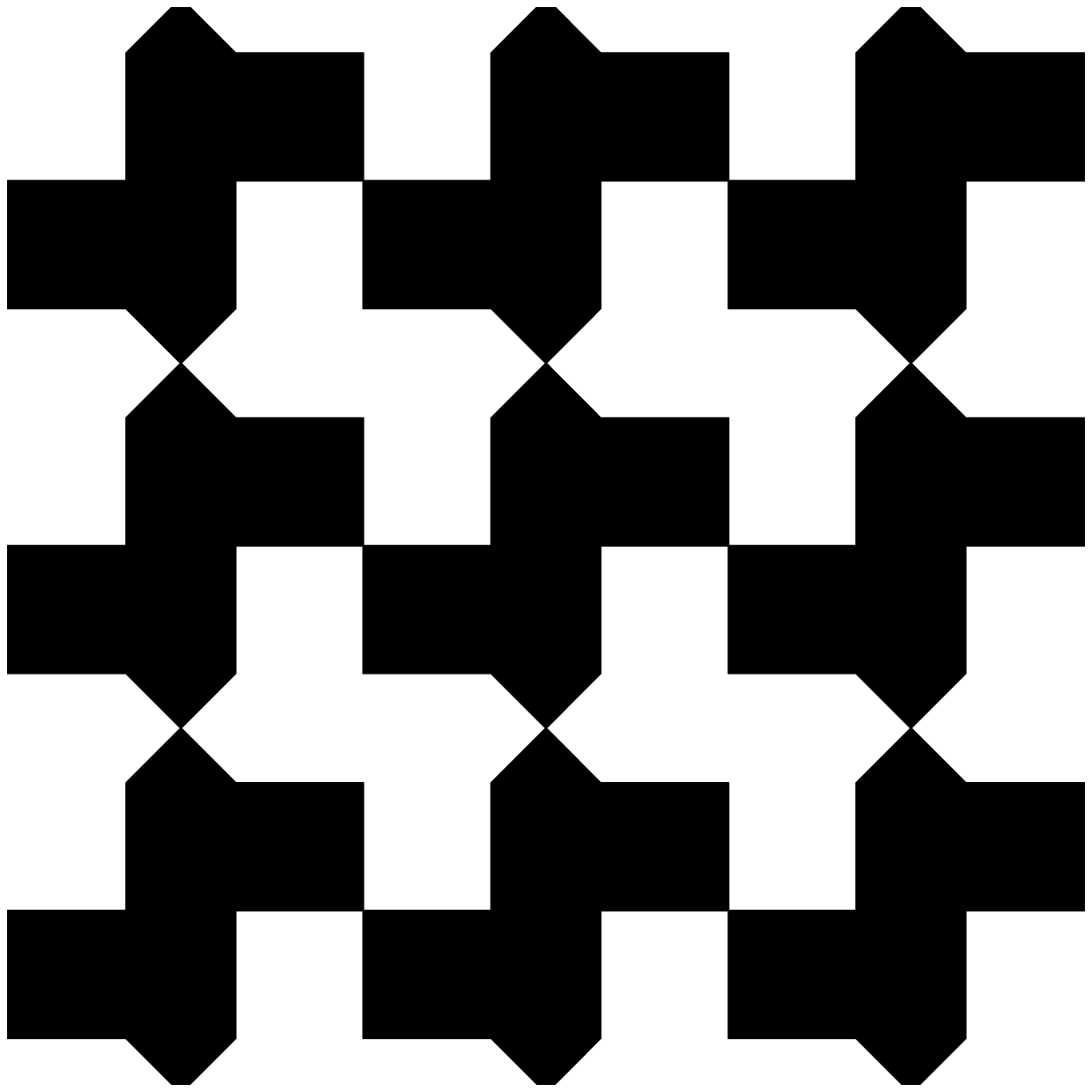}
\end{center}
\caption{Special textures arising in the theory: (a)
  stripes, which attain the Wiener bounds of maximal and
  minimal effective slip, if oriented parallel or perpendicular to the
  applied pressure gradient, respectively; (b) the Hashin-Shtrikman
  fractal pattern of circles, which attains the maximal slip among all
  isotropic textures; and (c) the Schulgasser texture, whose
  effective slip follows from the phase-interchange
  theorem.  }
\label{fig_designs}
\end{figure}

The upper bound for isotropic textures can be attained by a fractal
pattern of nested circular patches~\cite{Torquato:2002} as shown in
Fig~\ref{fig_designs}. It is interesting to note that similar patterns
might be expected for a random (and sometimes fractal) nanobubble
coating~\cite{vinogradova.oi:19952}.  However, it is not necessary to
deal with fractal surfaces: some periodic honeycomb-like structures
also attain the bound~\cite{Torquato-Gibiansky-Silva-Gibson:1998}.

Finally, we use phase interchange results~\cite{Torquato:2002} to
obtain the effective slip length without any calculations, for a
special class of isotropic textures.  For a medium that is invariant
by a $\pi/2$ rotation followed by a phase interchange, a classical
result follows: $ k^* = \sqrt{k_1 k_2}$. Examples of such media are
the chessboard and the Schulgasser texture, sketched in
Fig.~\ref{fig_designs}(c). The effective $\beta^*$ is then easily
obtained from (\ref{def_b_effective}), although the corresponding
values, shown in Fig.~\ref{fig_bounds}, are far from the HS upper
bound.

{\bf Concluding remarks: design strategies}-- We close by proposing
some guidelines for the design of thin superhydrophobic microchannels,
which maximize effective slippage, e.g. for lab-on-a-chip
applications. We assume a principal direction of the texture is
aligned with the side walls, since this is typically the fastest
orientation. (Tilted textures also complicate analysis, since the
constraint of no transverse flow, $\mean{U}_y=0$, induces a transverse
pressure gradient, $\mean{\nablab p}_y=-(k^*_{yx}/k^*_{yy})
\mean{\nablab p}_x$, which in turn affects the mean forward flow,
$\mean{U}_x=-(1/\mu)\, \mbox{det}(\kb^*)/k^*_{yy}$~\cite{tensor}.)
For simplicity, we also restrict now to the case $\beta_1=0$ of no-slip
support structures.

It has been predicted for thick ($L \ll h$)
cylindrical~\cite{lauga.e:2003} and planar channels
~\cite{wang2003,cottin.c:2004} that the longitudinal stripe
configuration has larger effective slip than the transverse one. For a
thin channel ($L \gg h$), we can now draw the more general conclusion
that longitudinal (transverse) stripes provide the largest (smallest)
possible slip that can be achieved by any texture. Interestingly, this
in contrast to a prediction for thick channels, where an array of
pillars in the limit $\phi_2 \to 1$ has larger slip than
longitudinal stripes~\cite{ybert2007}.

We have shown that the key parameter determining effective slip is the
area fraction of solid, $\phi_1$, in contact with the liquid. If this
is very small (or $\phi_2\to 1$), for all textures the effective slip
tends to a maximum, $\beta^*\to\beta_2$. In this limit, the
microchannel produces a kind of superfluidity, with plug-like
flow. However, even a very small $\phi_1$ is enough to reduce the
effective slip significantly since in this limit (except an upper
limit for case (II), where $\phi_2 - \beta^*/\beta_2 = 0$) we have the
asymptotic scaling $\phi_2 - \beta^*/\beta_2 \propto \beta_2
\phi_1$. It is interesting that in case of perfect slip over the gas
areas, $\beta^*$ scales as $\propto \phi_2/\phi_1$, which is similar
to an earlier result for a thick cylinder with transverse
stripes~\cite{lauga.e:2003}. For thin channels, we see now that this
result is very general and is valid for any texture (and likely any
channel geometry) with perfect slip patterns, representing
``obstacles'' to the flow. We thus conclude that in many situations,
maximizing $\beta_2$ is not nearly as important as optimizing the
texture to achieve large effective slip.

Finally, we have demonstrated that for all slip lengths and all
fractions the largest possible $\beta^*$ is equal to the area-averaged
slip length $\mean{\beta}$, attained by longitudinal superhydrophobic
stripes. However, for all textures the effective slip nearly coincide
with the average, provided $\beta_2$ is small (or, more generally,
$\beta_2-\beta_1$ is small). Although this limit is less important for
pressure-driven microfluidics, it may have relevance for amplifying
transport phenomena~\cite{ajdari2006}.

In summary, we have connected the problem of effective slip over
superhydrophobic surfaces in thin channels with the classical subject
of conduction in heterogeneous media. This has allowed us to obtain
rigorous bounds on slip for arbitrary textures and to obtain the slip
in some cases without any calculations. Our results can be used to
guide the design of superhydrophobic surfaces for thin micro- or
nano-channels (where slip is most important), and some principles may
hold for thick channels as well.

MZB and OIV gratefully acknowledge the hospitality of the ESPCI
through Paris-Sciences and Joliot Chairs. OIV was partly supported
by the DFG through its priority programme ``Micro- and
nanofluidics'' (grant Vi 243/1-3).

\bibliography{texture}

\begin{thebibliography}{32}
\expandafter\ifx\csname natexlab\endcsname\relax\def\natexlab#1{#1}\fi
\expandafter\ifx\csname bibnamefont\endcsname\relax
  \def\bibnamefont#1{#1}\fi
\expandafter\ifx\csname bibfnamefont\endcsname\relax
  \def\bibfnamefont#1{#1}\fi
\expandafter\ifx\csname citenamefont\endcsname\relax
  \def\citenamefont#1{#1}\fi
\expandafter\ifx\csname url\endcsname\relax
  \def\url#1{\texttt{#1}}\fi
\expandafter\ifx\csname urlprefix\endcsname\relax\def\urlprefix{URL }\fi
\providecommand{\bibinfo}[2]{#2}
\providecommand{\eprint}[2][]{\url{#2}}

\bibitem[{\citenamefont{Quere}(2005)}]{quere.d:2005}
\bibinfo{author}{\bibfnamefont{D.}~\bibnamefont{Quere}}, \bibinfo{journal}{Rep.
  Prog. Phys.} \textbf{\bibinfo{volume}{68}}, \bibinfo{pages}{2495}
  (\bibinfo{year}{2005}).

\bibitem[{\citenamefont{Bico et~al.}(2002)\citenamefont{Bico, Thiele, and
  Quere}}]{bico.j:2002}
\bibinfo{author}{\bibfnamefont{J.}~\bibnamefont{Bico}},
  \bibinfo{author}{\bibfnamefont{U.}~\bibnamefont{Thiele}}, \bibnamefont{and}
  \bibinfo{author}{\bibfnamefont{D.}~\bibnamefont{Quere}},
  \bibinfo{journal}{Colloids Surfaces A} \textbf{\bibinfo{volume}{206}},
  \bibinfo{pages}{41} (\bibinfo{year}{2002}).

\bibitem[{\citenamefont{{Stone} et~al.}(2004)\citenamefont{{Stone}, {Stroock},
  and {Ajdari}}}]{stone2004}
\bibinfo{author}{\bibfnamefont{H.~A.} \bibnamefont{{Stone}}},
  \bibinfo{author}{\bibfnamefont{A.~D.} \bibnamefont{{Stroock}}},
  \bibnamefont{and} \bibinfo{author}{\bibfnamefont{A.}~\bibnamefont{{Ajdari}}},
  \bibinfo{journal}{Annual Review of Fluid Mechanics}
  \textbf{\bibinfo{volume}{36}}, \bibinfo{pages}{381} (\bibinfo{year}{2004}).

\bibitem[{\citenamefont{Squires and Quake}(2005)}]{squires2005}
\bibinfo{author}{\bibfnamefont{T.~M.} \bibnamefont{Squires}} \bibnamefont{and}
  \bibinfo{author}{\bibfnamefont{S.~R.} \bibnamefont{Quake}},
  \bibinfo{journal}{Reviews of Modern Physics} \textbf{\bibinfo{volume}{77}},
  \bibinfo{pages}{977} (\bibinfo{year}{2005}).

\bibitem[{\citenamefont{Ajdari and Bocquet}(2006)}]{ajdari2006}
\bibinfo{author}{\bibfnamefont{A.}~\bibnamefont{Ajdari}} \bibnamefont{and}
  \bibinfo{author}{\bibfnamefont{L.}~\bibnamefont{Bocquet}},
  \bibinfo{journal}{Phys. Rev. Lett.} \textbf{\bibinfo{volume}{96}},
  \bibinfo{pages}{186102} (\bibinfo{year}{2006}).

\bibitem[{\citenamefont{{Stroock} et~al.}(2002)\citenamefont{{Stroock},
  {Dertinger}, {Ajdari}, {Mezi{\'c}}, {Stone}, and
  {Whitesides}}}]{stroock2002a}
\bibinfo{author}{\bibfnamefont{A.~D.} \bibnamefont{{Stroock}}},
  \bibinfo{author}{\bibfnamefont{S.~K.~W.} \bibnamefont{{Dertinger}}},
  \bibinfo{author}{\bibfnamefont{A.}~\bibnamefont{{Ajdari}}},
  \bibinfo{author}{\bibfnamefont{I.}~\bibnamefont{{Mezi{\'c}}}},
  \bibinfo{author}{\bibfnamefont{H.~A.} \bibnamefont{{Stone}}},
  \bibnamefont{and} \bibinfo{author}{\bibfnamefont{G.~M.}
  \bibnamefont{{Whitesides}}}, \bibinfo{journal}{Science}
  \textbf{\bibinfo{volume}{295}}, \bibinfo{pages}{647} (\bibinfo{year}{2002}).

\bibitem[{\citenamefont{Vinogradova}(1999)}]{vinogradova.oi:1999}
\bibinfo{author}{\bibfnamefont{O.~I.} \bibnamefont{Vinogradova}},
  \bibinfo{journal}{Int. J. Mineral Proces.} \textbf{\bibinfo{volume}{56}},
  \bibinfo{pages}{31} (\bibinfo{year}{1999}).

\bibitem[{\citenamefont{Lauga et~al.}(2007)\citenamefont{Lauga, Brenner, and
  Stone}}]{lauga2007}
\bibinfo{author}{\bibfnamefont{E.}~\bibnamefont{Lauga}},
  \bibinfo{author}{\bibfnamefont{M.~P.} \bibnamefont{Brenner}},
  \bibnamefont{and} \bibinfo{author}{\bibfnamefont{H.~A.} \bibnamefont{Stone}},
  \emph{\bibinfo{title}{Handbook of Experimental Fluid Dynamics}}
  (\bibinfo{publisher}{Springer}, \bibinfo{address}{NY}, \bibinfo{year}{2007}),
  chap.~\bibinfo{chapter}{19}, pp. \bibinfo{pages}{1219--1240}.

\bibitem[{\citenamefont{{Bocquet} and Barrat}(2007)}]{bocquet2007}
\bibinfo{author}{\bibfnamefont{L.}~\bibnamefont{{Bocquet}}} \bibnamefont{and}
  \bibinfo{author}{\bibfnamefont{J.~L.} \bibnamefont{Barrat}},
  \bibinfo{journal}{Soft Matter} \textbf{\bibinfo{volume}{3}},
  \bibinfo{pages}{685} (\bibinfo{year}{2007}).

\bibitem[{\citenamefont{Vinogradova}(1995)}]{vinogradova.oi:19951}
\bibinfo{author}{\bibfnamefont{O.~I.} \bibnamefont{Vinogradova}},
  \bibinfo{journal}{Langmuir} \textbf{\bibinfo{volume}{11}},
  \bibinfo{pages}{2213} (\bibinfo{year}{1995}).

\bibitem[{not({\natexlab{a}})}]{note0}
\bibinfo{note}{A variant of this picture is a nanobubble-coated hydrophobic
  surface~\cite{vinogradova.oi:19952,borkent.bm:2007}.}

\bibitem[{\citenamefont{Vinogradova and Yakubov}(2003)}]{vinogradova.oi:2003}
\bibinfo{author}{\bibfnamefont{O.~I.} \bibnamefont{Vinogradova}}
  \bibnamefont{and} \bibinfo{author}{\bibfnamefont{G.~E.}
  \bibnamefont{Yakubov}}, \bibinfo{journal}{Langmuir}
  \textbf{\bibinfo{volume}{19}}, \bibinfo{pages}{1227} (\bibinfo{year}{2003}).

\bibitem[{\citenamefont{Bazant and Vinogradova}(2008)}]{tensor}
\bibinfo{author}{\bibfnamefont{M.~Z.} \bibnamefont{Bazant}} \bibnamefont{and}
  \bibinfo{author}{\bibfnamefont{O.~I.} \bibnamefont{Vinogradova}},
  \bibinfo{journal}{J. Fluid Mech.}  (\bibinfo{year}{2008}), \bibinfo{note}{in
  press}.

\bibitem[{\citenamefont{Joseph et~al.}(2006)\citenamefont{Joseph, Cottin-Bizon,
  Benoit, Ybert, Journet, Tabeling, and Bocquet}}]{joseph.p:2006}
\bibinfo{author}{\bibfnamefont{P.}~\bibnamefont{Joseph}},
  \bibinfo{author}{\bibfnamefont{C.}~\bibnamefont{Cottin-Bizon}},
  \bibinfo{author}{\bibfnamefont{J.~M.} \bibnamefont{Benoit}},
  \bibinfo{author}{\bibfnamefont{C.}~\bibnamefont{Ybert}},
  \bibinfo{author}{\bibfnamefont{C.}~\bibnamefont{Journet}},
  \bibinfo{author}{\bibfnamefont{P.}~\bibnamefont{Tabeling}}, \bibnamefont{and}
  \bibinfo{author}{\bibfnamefont{L.}~\bibnamefont{Bocquet}},
  \bibinfo{journal}{Phys. Rev. Lett.} \textbf{\bibinfo{volume}{96}},
  \bibinfo{pages}{156104} (\bibinfo{year}{2006}).

\bibitem[{\citenamefont{Ou and Rothstein}(2005)}]{ou.j:2005}
\bibinfo{author}{\bibfnamefont{J.}~\bibnamefont{Ou}} \bibnamefont{and}
  \bibinfo{author}{\bibfnamefont{J.~P.} \bibnamefont{Rothstein}},
  \bibinfo{journal}{Phys. Fluids} \textbf{\bibinfo{volume}{17}},
  \bibinfo{pages}{103606} (\bibinfo{year}{2005}).

\bibitem[{\citenamefont{Lauga and Stone}(2003)}]{lauga.e:2003}
\bibinfo{author}{\bibfnamefont{E.}~\bibnamefont{Lauga}} \bibnamefont{and}
  \bibinfo{author}{\bibfnamefont{H.~A.} \bibnamefont{Stone}},
  \bibinfo{journal}{J. Fluid Mech.} \textbf{\bibinfo{volume}{489}},
  \bibinfo{pages}{55} (\bibinfo{year}{2003}).

\bibitem[{\citenamefont{{Wang}}(2003)}]{wang2003}
\bibinfo{author}{\bibfnamefont{C.~Y.} \bibnamefont{{Wang}}},
  \bibinfo{journal}{Physics of Fluids} \textbf{\bibinfo{volume}{15}},
  \bibinfo{pages}{1114} (\bibinfo{year}{2003}).

\bibitem[{\citenamefont{Sbragaglia and Prosperetti}(2007)}]{sbragaglia.m:2007}
\bibinfo{author}{\bibfnamefont{M.}~\bibnamefont{Sbragaglia}} \bibnamefont{and}
  \bibinfo{author}{\bibfnamefont{A.}~\bibnamefont{Prosperetti}},
  \bibinfo{journal}{Phys. Fluids} \textbf{\bibinfo{volume}{19}},
  \bibinfo{pages}{043603} (\bibinfo{year}{2007}).

\bibitem[{\citenamefont{Alexeyev and Vinogradova}(1996)}]{alexeyev:96}
\bibinfo{author}{\bibfnamefont{A.~A.} \bibnamefont{Alexeyev}} \bibnamefont{and}
  \bibinfo{author}{\bibfnamefont{O.~I.} \bibnamefont{Vinogradova}},
  \bibinfo{journal}{Colloids Surfaces A} \textbf{\bibinfo{volume}{108}},
  \bibinfo{pages}{173 } (\bibinfo{year}{1996}).

\bibitem[{\citenamefont{{Ybert} et~al.}(2007)\citenamefont{{Ybert}, {Barentin},
  {Cottin-Bizonne}, {Joseph}, and {Bocquet}}}]{ybert2007}
\bibinfo{author}{\bibfnamefont{C.}~\bibnamefont{{Ybert}}},
  \bibinfo{author}{\bibfnamefont{C.}~\bibnamefont{{Barentin}}},
  \bibinfo{author}{\bibfnamefont{C.}~\bibnamefont{{Cottin-Bizonne}}},
  \bibinfo{author}{\bibfnamefont{P.}~\bibnamefont{{Joseph}}}, \bibnamefont{and}
  \bibinfo{author}{\bibfnamefont{L.}~\bibnamefont{{Bocquet}}},
  \bibinfo{journal}{Physics of Fluids} \textbf{\bibinfo{volume}{19}},
  \bibinfo{pages}{123601} (\bibinfo{year}{2007}).

\bibitem[{\citenamefont{Cottin-Bizonne
  et~al.}(2004)\citenamefont{Cottin-Bizonne, Barentin, Charlaix, Bocquet, and
  Barrat}}]{cottin.c:2004}
\bibinfo{author}{\bibfnamefont{C.}~\bibnamefont{Cottin-Bizonne}},
  \bibinfo{author}{\bibfnamefont{C.}~\bibnamefont{Barentin}},
  \bibinfo{author}{\bibfnamefont{E.}~\bibnamefont{Charlaix}},
  \bibinfo{author}{\bibfnamefont{L.}~\bibnamefont{Bocquet}}, \bibnamefont{and}
  \bibinfo{author}{\bibfnamefont{J.~L.} \bibnamefont{Barrat}},
  \bibinfo{journal}{Eur. Phys. J. E} \textbf{\bibinfo{volume}{15}},
  \bibinfo{pages}{427} (\bibinfo{year}{2004}).

\bibitem[{\citenamefont{Priezjev et~al.}(2005)\citenamefont{Priezjev, Darhuber,
  and Troian}}]{priezjev.nv:2005}
\bibinfo{author}{\bibfnamefont{N.~V.} \bibnamefont{Priezjev}},
  \bibinfo{author}{\bibfnamefont{A.~A.} \bibnamefont{Darhuber}},
  \bibnamefont{and} \bibinfo{author}{\bibfnamefont{S.~M.}
  \bibnamefont{Troian}}, \bibinfo{journal}{Phys. Rev. E}
  \textbf{\bibinfo{volume}{71}}, \bibinfo{pages}{041608}
  (\bibinfo{year}{2005}).

\bibitem[{\citenamefont{Benzi et~al.}(2006)\citenamefont{Benzi, Biferale,
  Sbragaglia, Succi, and Toschi}}]{benzi.r:2006}
\bibinfo{author}{\bibfnamefont{R.}~\bibnamefont{Benzi}},
  \bibinfo{author}{\bibfnamefont{L.}~\bibnamefont{Biferale}},
  \bibinfo{author}{\bibfnamefont{M.}~\bibnamefont{Sbragaglia}},
  \bibinfo{author}{\bibfnamefont{S.}~\bibnamefont{Succi}}, \bibnamefont{and}
  \bibinfo{author}{\bibfnamefont{F.}~\bibnamefont{Toschi}},
  \bibinfo{journal}{J. Fluid Mech.} \textbf{\bibinfo{volume}{548}},
  \bibinfo{pages}{257} (\bibinfo{year}{2006}).

\bibitem[{\citenamefont{Markov}(2000)}]{Markov:2000}
\bibinfo{author}{\bibfnamefont{K.~Z.} \bibnamefont{Markov}}, in
  \emph{\bibinfo{booktitle}{Heterogeneous Media, Modelling and Simulation}},
  edited by \bibinfo{editor}{\bibfnamefont{K.}~\bibnamefont{Markov}}
  \bibnamefont{and} \bibinfo{editor}{\bibfnamefont{L.}~\bibnamefont{Preziosi}}
  (\bibinfo{publisher}{Birkhauser Boston}, \bibinfo{year}{2000}),
  chap.~\bibinfo{chapter}{1}, pp. \bibinfo{pages}{1--162}.

\bibitem[{\citenamefont{Torquato}(2002)}]{Torquato:2002}
\bibinfo{author}{\bibfnamefont{S.}~\bibnamefont{Torquato}},
  \emph{\bibinfo{title}{Random Heterogeneous Materials: Microstructure and
  Macroscopic Properties}} (\bibinfo{publisher}{Springer},
  \bibinfo{year}{2002}).

\bibitem[{\citenamefont{Hyv\"aluoma and Harting}(2008)}]{hyvaluoma.j:2008}
\bibinfo{author}{\bibfnamefont{J.}~\bibnamefont{Hyv\"aluoma}} \bibnamefont{and}
  \bibinfo{author}{\bibfnamefont{J.}~\bibnamefont{Harting}},
  \bibinfo{journal}{Phys. Rev. Lett.} \textbf{\bibinfo{volume}{100}},
  \bibinfo{pages}{246001} (\bibinfo{year}{2008}).

\bibitem[{not({\natexlab{b}})}]{note2}
\bibinfo{note}{This raises interesting questions about correlations between an
  effective slip and static wetting properties (contact angle) of the
  wall~\cite{voronov.rs:2008}. For example, in the stripe geometry the
  ``transverse'' contact angle is larger than ``parallel''~\cite{bico.j:1999},
  which is opposite to trends in slippage.}

\bibitem[{\citenamefont{Vinogradova et~al.}(1995)\citenamefont{Vinogradova,
  Bunkin, Churaev, Kiseleva, and Ninham}}]{vinogradova.oi:19952}
\bibinfo{author}{\bibfnamefont{O.~I.} \bibnamefont{Vinogradova}},
  \bibinfo{author}{\bibfnamefont{N.~F.} \bibnamefont{Bunkin}},
  \bibinfo{author}{\bibfnamefont{N.~V.} \bibnamefont{Churaev}},
  \bibinfo{author}{\bibfnamefont{O.~A.} \bibnamefont{Kiseleva}},
  \bibnamefont{and} \bibinfo{author}{\bibfnamefont{B.~W.}
  \bibnamefont{Ninham}}, \bibinfo{journal}{J. Colloid Interface Sci.}
  \textbf{\bibinfo{volume}{173}}, \bibinfo{pages}{443} (\bibinfo{year}{1995}).

\bibitem[{\citenamefont{Torquato et~al.}(1998)\citenamefont{Torquato,
  Gibiansky, Silva, and Gibson}}]{Torquato-Gibiansky-Silva-Gibson:1998}
\bibinfo{author}{\bibfnamefont{S.}~\bibnamefont{Torquato}},
  \bibinfo{author}{\bibfnamefont{L.}~\bibnamefont{Gibiansky}},
  \bibinfo{author}{\bibfnamefont{M.}~\bibnamefont{Silva}}, \bibnamefont{and}
  \bibinfo{author}{\bibfnamefont{L.}~\bibnamefont{Gibson}},
  \bibinfo{journal}{Int. J. Mech. Sci.} \textbf{\bibinfo{volume}{40}},
  \bibinfo{pages}{71} (\bibinfo{year}{1998}).

\bibitem[{\citenamefont{Borkent et~al.}(2007)\citenamefont{Borkent, Dammler,
  Sch\"onherr, Vansco, and Lohse}}]{borkent.bm:2007}
\bibinfo{author}{\bibfnamefont{B.~M.} \bibnamefont{Borkent}},
  \bibinfo{author}{\bibfnamefont{S.~M.} \bibnamefont{Dammler}},
  \bibinfo{author}{\bibfnamefont{H.}~\bibnamefont{Sch\"onherr}},
  \bibinfo{author}{\bibfnamefont{G.~J.} \bibnamefont{Vansco}},
  \bibnamefont{and} \bibinfo{author}{\bibfnamefont{D.}~\bibnamefont{Lohse}},
  \bibinfo{journal}{Phys. Rev. Lett.} \textbf{\bibinfo{volume}{98}},
  \bibinfo{pages}{204502} (\bibinfo{year}{2007}).

\bibitem[{\citenamefont{Voronov et~al.}(2008)\citenamefont{Voronov,
  Papavassiliou, and Lee}}]{voronov.rs:2008}
\bibinfo{author}{\bibfnamefont{R.~S.} \bibnamefont{Voronov}},
  \bibinfo{author}{\bibfnamefont{D.~V.} \bibnamefont{Papavassiliou}},
  \bibnamefont{and} \bibinfo{author}{\bibfnamefont{L.~L.} \bibnamefont{Lee}},
  \bibinfo{journal}{Ind. Eng. Chem. Res.} \textbf{\bibinfo{volume}{47}},
  \bibinfo{pages}{2455} (\bibinfo{year}{2008}).

\bibitem[{\citenamefont{Bico et~al.}(1999)\citenamefont{Bico, Marzolin, and
  Quere}}]{bico.j:1999}
\bibinfo{author}{\bibfnamefont{J.}~\bibnamefont{Bico}},
  \bibinfo{author}{\bibfnamefont{C.}~\bibnamefont{Marzolin}}, \bibnamefont{and}
  \bibinfo{author}{\bibfnamefont{D.}~\bibnamefont{Quere}},
  \bibinfo{journal}{Europhys. Lett.} \textbf{\bibinfo{volume}{47}},
  \bibinfo{pages}{220} (\bibinfo{year}{1999}).

\end{thebibliography}
\end{document}